\documentclass[prl,twocolumn,groupedaddress,showpacs,byrevtex]{revtex4}
\usepackage[]{umlaut}
\usepackage[]{amssymb}
\usepackage[]{graphicx} 
\usepackage[]{dcolumn}

\begin{document}

\bibliographystyle{prlsty}

\title{The Metal-Insulator Transition of $ \rm {\bf NbO_2} $: 
       an Embedded Peierls Instability}

\author{V.\ Eyert} 
\homepage[http:]{//www.physik.uni-augsburg.de/~eyert/}
\affiliation{Institut f\"ur Physik, Universit\"at Augsburg, 
             86135 Augsburg, Germany}
\date{\today}

\begin{abstract}
Results of first principles augmented spherical wave electronic structure 
calculations for niobium dioxide are presented. Both metallic rutile and 
insulating low-temperature $ {\rm NbO_2} $, which crystallizes in a 
distorted rutile structure, are correctly described within density functional 
theory and the local density approximation. Metallic conductivity is 
carried to equal amounts by metal $ t_{2g } $ orbitals, which fall into 
the one-dimensional $ d_{\parallel} $ band and the isotropically dispersing 
$ e_{g}^{\pi} $ bands. Hybridization of both types of bands is almost 
negligible outside narrow rods along the line X--R. In the 
low-temperature phase splitting of the $ d_{\parallel} $ band due to 
metal-metal dimerization as well as upshift of the $ e_{g}^{\pi} $ bands 
due to increased $ p $-$ d $ overlap remove the Fermi surface and open an 
optical band gap of about 0.1 eV. The metal-insulator transition arises 
as a Peierls instability of the $ d_{\parallel} $ band in an embedding 
background of $ e_{g}^{\pi} $ electrons. This basic mechanism should 
also apply to $ {\rm VO_2} $, where, however, electronic correlations 
are expected to play a greater role due to stronger localization of the 
$ 3d $ electrons.  
\end{abstract}


\maketitle

Despite intense work over decades the metal-insulator transition (MIT) of 
$ {\rm VO_2} $ has remained a matter of controversy 
\cite{goode60,adler68,goode71b,zylbersztejn75,wentz94,rice94,imada98}. 
This is related to the simultaneous occurrence of a structural 
transformation from the high-temperature rutile phase to a distorted 
monoclinic structure, which is characterized by (i) pairing of the 
metal atoms within chains parallel to the rutile $ c $ axis and (ii) their 
lateral zigzag-like displacement \cite{andersson56longo70}. 
Electronic states near the Fermi energy are of mainly V $ 3d $ $ t_{2g} $ 
character. They separate into the $ d_{\parallel} $ band, 
which mediates V-V overlap along the metal chains, and the remaining 
$ e_g^{\pi} $ bands \cite{goode71b}. At the transition, 
splitting of the $ d_{\parallel} $ band and upshift of the $ e_g^{\pi} $ 
bands due to increased metal-oxygen overlap produce a finite band 
gap. Dispute centers about the question, whether the $ d_{\parallel} $ 
splitting is caused by metal dimerization or by increased electronic 
correlations resulting from the reduced screening by the $ e_g^{\pi} $ 
electrons \cite{goode71b,zylbersztejn75}. State of the art band 
calculations gave strong hints of a structural instability but were 
not able to reproduce the insulating gap due to the shortcomings of the 
local density approximation (LDA) \cite{wentz94}. 

Interestingly, only very few studies have dealt with the neighbouring 
dioxides, which display related phenomena. $ {\rm NbO_2} $ likewise 
undergoes a MIT (at 1081 K) and a simultaneous structural transition 
from rutile to a distorted variant having a body-centered tetragonal 
(bct) lattice \cite{janninck66,sakata67}. Despite the differences in 
long range order, local deviations from rutile are the same as in 
$ {\rm VO_2} $, i.e.\ niobium atoms dimerize and experience lateral 
displacements at the transition \cite{marinder62,cheetham76,pynn76b}. 
In contrast, the metallic oxides $ {\rm MoO_2} $, $ {\rm WO_2} $, 
$ {\rm TcO_2} $, and $ \alpha $-$ {\rm ReO_2} $ all crystallize in the 
same monoclinic structure as $ {\rm VO_2} $ \cite{rogers69mattheiss76}. 
No doubt the phase transitions of $ {\rm VO_2} $ and $ {\rm NbO_2} $ as 
well as the destabilization of the rutile structure in all these dioxides 
call for a unified description \cite{sakata67,adler68,goode71b}. Yet, 
although work on the dioxides neighbouring $ {\rm VO_2} $ would help 
clarifying the nature of the MIT, there are only few theoretical 
investigations in this direction \cite{posternak79,sasaki81,habilmoo2pap}. 
For this reason, a complete and widely accepted picture of the 
rutile-related transition metal dioxides has not yet evolved. 

Here we present first principles electronic structure calculations for 
$ {\rm NbO_2} $. These are the first calculations at all for the 
insulating phase and the first calculations for rutile $ {\rm NbO_2} $, 
which used the experimental crystal structure. Both phases are correctly 
described within density functional theory (DFT) and the LDA. The results 
strongly support the band theoretical point of view of the MIT as arising 
predominantly from a Peierls-type instability of the rutile phase. 

Evidence for the MIT being driven by strong metal-metal bonding was first 
inferred from the small and almost temperature independent paramagnetic 
susceptibility below the transition temperature as well as from the 
enhanced thermal expansion of the $ c $ lattice parameter just above 
$ T_c $ \cite{brauer41,ruedorff64,sakata69,sakata69d}. Additional support 
came from strong decrease of $ T_c $, nonlinear variation of $ c $, and 
crossover of the magnetic susceptibility to Curie-Weiss behaviour on 
substitution of Ti for Nb, which destroys the metal-metal bonding 
\cite{ruedorff64,sakata69,sakata69d}. 

According to neutron-diffraction experiments the bct structure of 
insulating $ {\rm NbO_2} $ has 16 formula units per cell, space 
group $ I4_1/a $ ($ C_{4h}^{6} $), and lattice constants 
$ a \approx 2 \sqrt{2} a_R $ and $ c \approx 2 c_R $ 
\cite{marinder62,cheetham76,pynn76b}, where $ a_R $ = 4.8464{\AA} and 
$ c_R $ = 3.0316{\AA} are the rutile lattice parameters \cite{bolzan94}.  
The two different Nb-Nb distances of 2.7 and 3.3{\AA} as arising from 
the metal-metal pairing along the rutile $ c $ axis \cite{pynn76b} deviate 
considerably from the rutile value $ c_R $. In addition, the zigzag-like 
in-plane displacement of niobium atoms parallel to either the $ [110] $ 
or $ [1\bar{1}0] $ direction of the rutile subcell leads to a variation 
of Nb--O distances between 1.91 and 2.25{\AA}, which deviate from the  
high-temperature values of 2.00 and 2.08{\AA} \cite{cheetham76,bolzan94}. 
Critical scattering observed above $ T_c $ at the tetragonal wave vector 
$ {\bf q}_P = (1/4,1/4,1/2) $ pointed to a soft phonon mode, which, 
however, could not be clearly identified \cite{shapiro74,pynn76b,pynn78}. 
Yet, using a shell model, Gervais and Kress were able to reveal softening 
of a $ P $ point phonon with a displacement pattern consistent with the 
low-temperature structure \cite{gervais85}. 

Room-temperature UPS and XPS experiments revealed a 9 eV wide occupied 
group of bands, which falls into the 1 eV wide Nb $ 4d $ bands just below 
$ E_F $ and a 6 eV wide group of O $ 2p $ bands at higher binding energies 
\cite{beatham79}. The room temperature optical band gap was estimated to 
about 0.5 eV \cite{adler68}. By and large, photoemission data agree with 
band structure and X$ \alpha $ cluster calculations 
\cite{posternak79,umrigar80,sasaki81,xu89}, which, due to the complexitiy 
of the low-temperature structure, only addressed to the rutile 
phase.  Only Sasaki {\em et al.}\ accounted for the bct structure by 
studying $ {\rm Nb_2O_{10}} $ clusters with either short or long Nb-Nb 
distance and deduced a band gap of 0.68 eV; however, oxygen 
positions were not relaxed and the same value for all Nb-O distances 
used \cite{sasaki81}. Finally, calculated Fermi surfaces and generalized 
susceptibilities agreed with experiment in that they could not find a 
soft-mode instability \cite{posternak79,xu89}. 

The present calculations were performed using the scalar-relativistic 
augmented spherical wave (ASW) method \cite{wkgrevasw}. Both the LDA 
and the generalized gradient approximation were applied with essentially 
no differences in the electronic structure. Crystallographic data as 
given by Bolzan {\em et al.}\ \cite{bolzan94} and Pynn {\em et al.}\ 
\cite{pynn76b} were used.  In order to account for the loose packing of 
the crystal structures empty spheres, i.e.\ pseudo atoms without a nucleus, 
were included to model the correct shape of the crystal potential in large 
voids. Optimal empty sphere positions and radii of all spheres were 
automatically determined \cite{vpop}. As a result, three and 14 
in-equivalent empty sphere types with radii ranging from 0.96 to 2.49 
$ a_B $ were included in the rutile and bct cell, respectively, keeping 
the linear overlap of niobium and oxygen spheres below 18.5\%. The basis 
set comprised Nb $ 5s $, $ 5p $, $ 4d $ and O $ 2s $, $ 2p $ as well as 
empty sphere states. Brillouin zone sampling was done using an increased 
number of $ {\bf k} $ points ranging from 18 to 1800 and 12 to 552 points  
within the respective irreducible wedge. 

\begin{figure}[htb]
  \begin{center}
    \includegraphics[width=8.0cm]{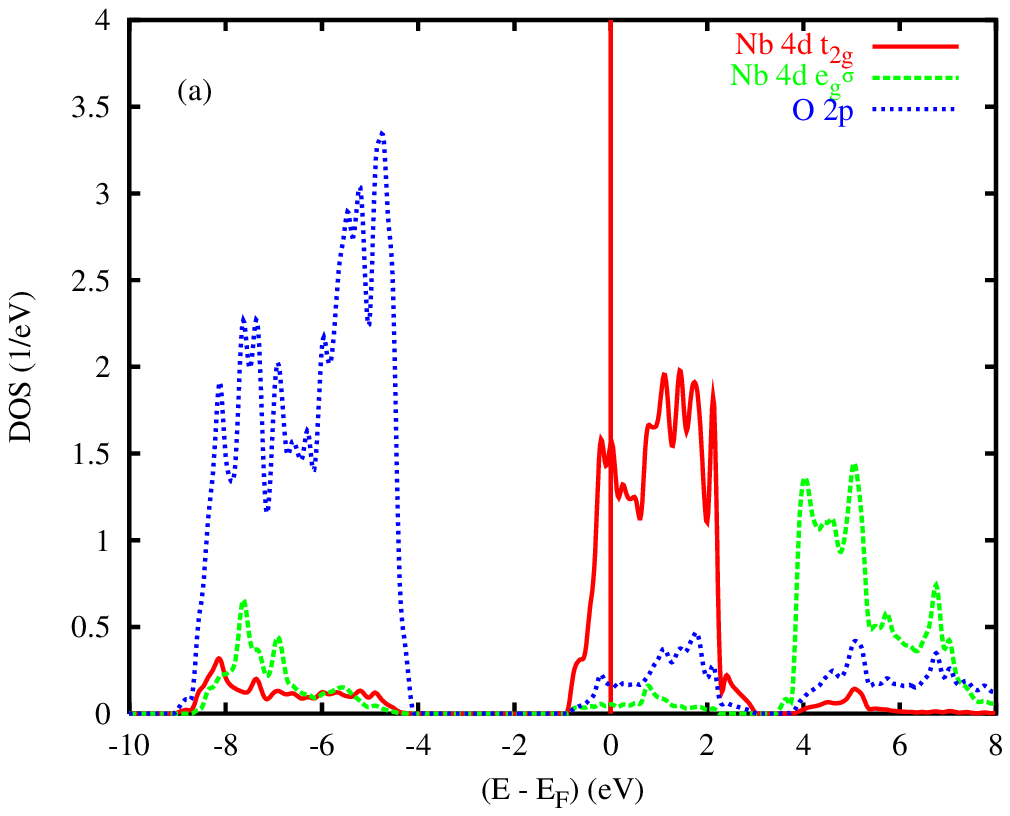}
    \includegraphics[width=8.0cm]{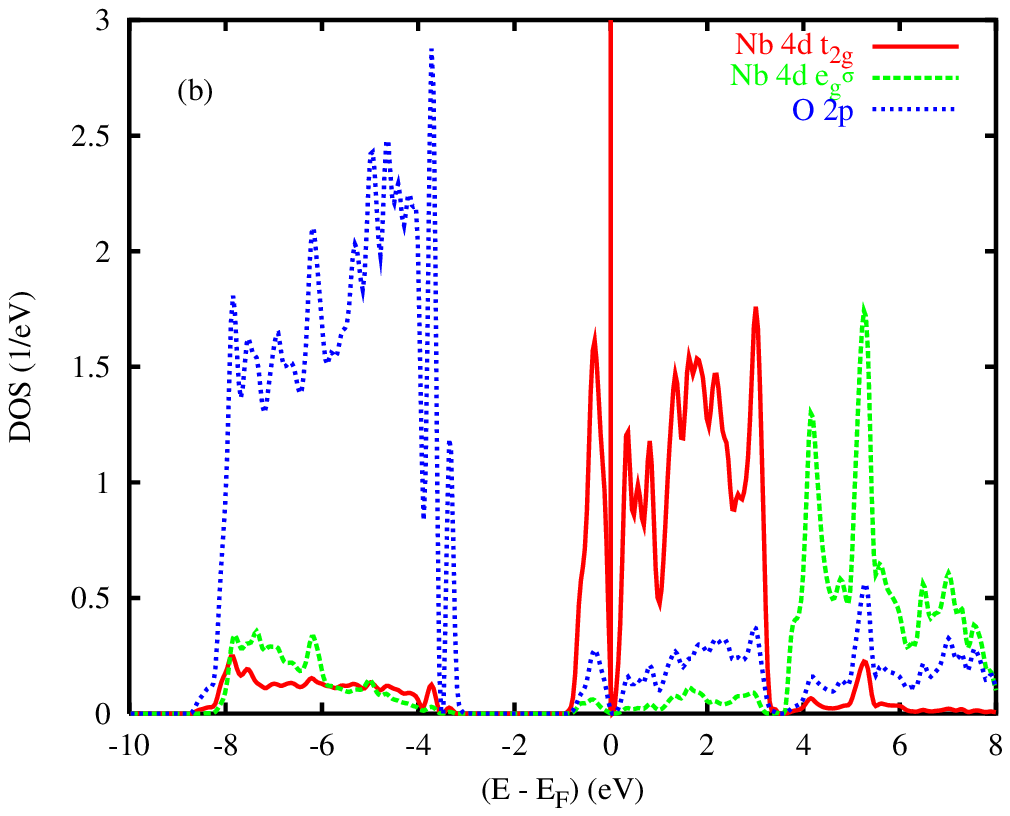}
    \caption{Partial DOS of (a) metallic and (b) insulating 
             $ {\rm NbO_2} $ (per f.u.). 
             Slight broadening is due to the DOS calculation scheme 
             \protect \cite{methfessel89}.}
    \label{fig:dosrut}
  \end{center}
\end{figure}

Partial densities of states (DOS) resulting from calculations for the 
rutile and bct structures are displayed in Fig.\ \ref{fig:dosrut}. 
The gross features are similar in both phases. Three groups 
of bands are identified. While bands in the energy range from $ \approx -9 $ 
to $ -3 $ eV are dominated by O $ 2p $ states, the two higher lying groups 
at and above the Fermi energy derive mainly from Nb $ 4d $ orbitals. All 
other states play only a negligible role in the energy interval shown. 
Crystal field splitting causes nearly complete energetical separation of 
the Nb $ 4d $ bands into two groups of $ t_{2g} $ and $ e_g^{\sigma} $ 
symmetry. Contributions of the O $ 2p $ and Nb $ 4d $ states to the upper 
two and the lower group, respectively, are indicative of covalent bonding. 

Our results for insulating and metallic $ {\rm NbO_2} $ are in good 
agreement with the photoemission experiments \cite{beatham79} and 
the previous calculations \cite{posternak79,umrigar80,sasaki81,xu89}. 
Differences with the latter could be traced back to the fact that these 
calculations were not self-consistent or, due to lack of crystallographic 
data, used a rutile structure obtained from symmetrizing the bct structure.  

\begin{figure}[htb]
  \begin{center}
    \includegraphics[width = 8.0cm]{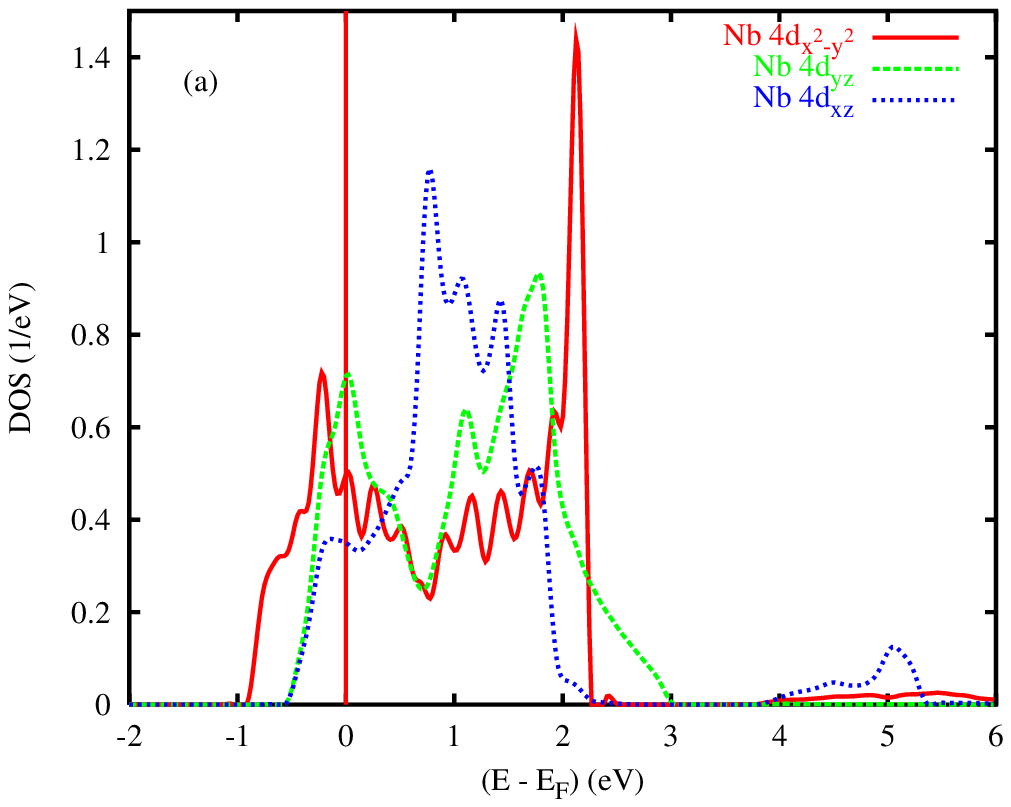}
    \includegraphics[width = 8.0cm]{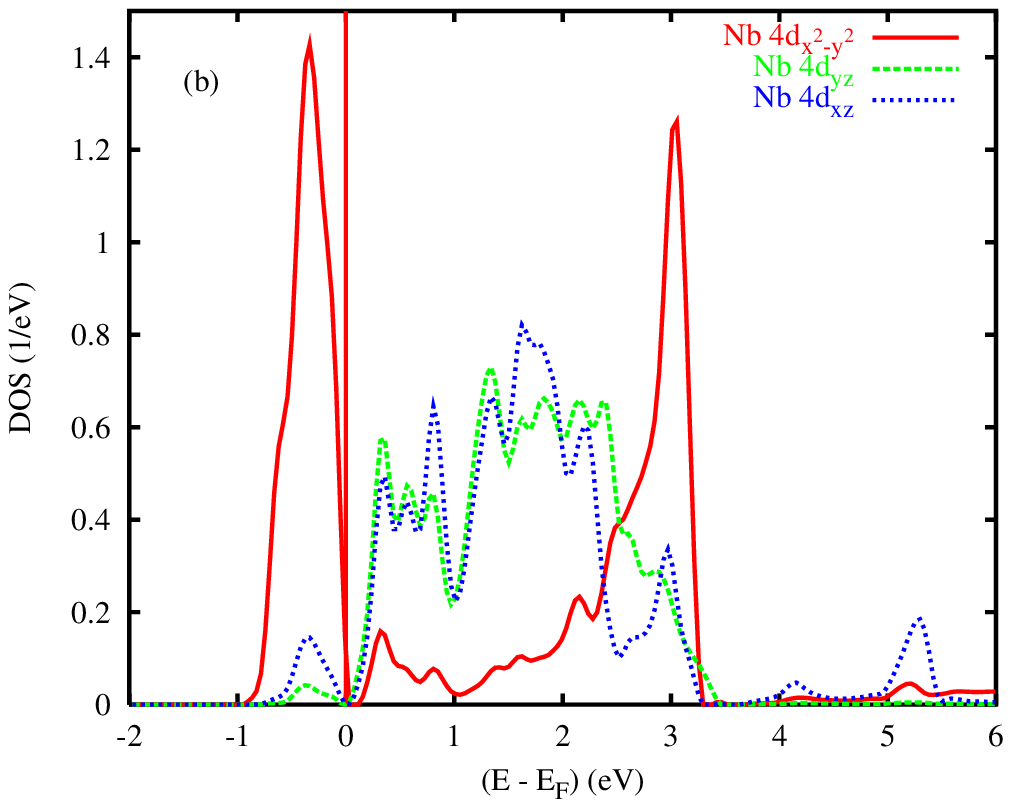}
    \caption{Partial $ t_{2g} $ DOS of (a) metallic and (b) insulating 
             $ {\rm NbO_2} $ (per f.u.).}
    \label{fig:dosrutt2g}
  \end{center}
\end{figure}

\begin{figure}[htb]
  \begin{center}
    \includegraphics[width = 8.0cm]{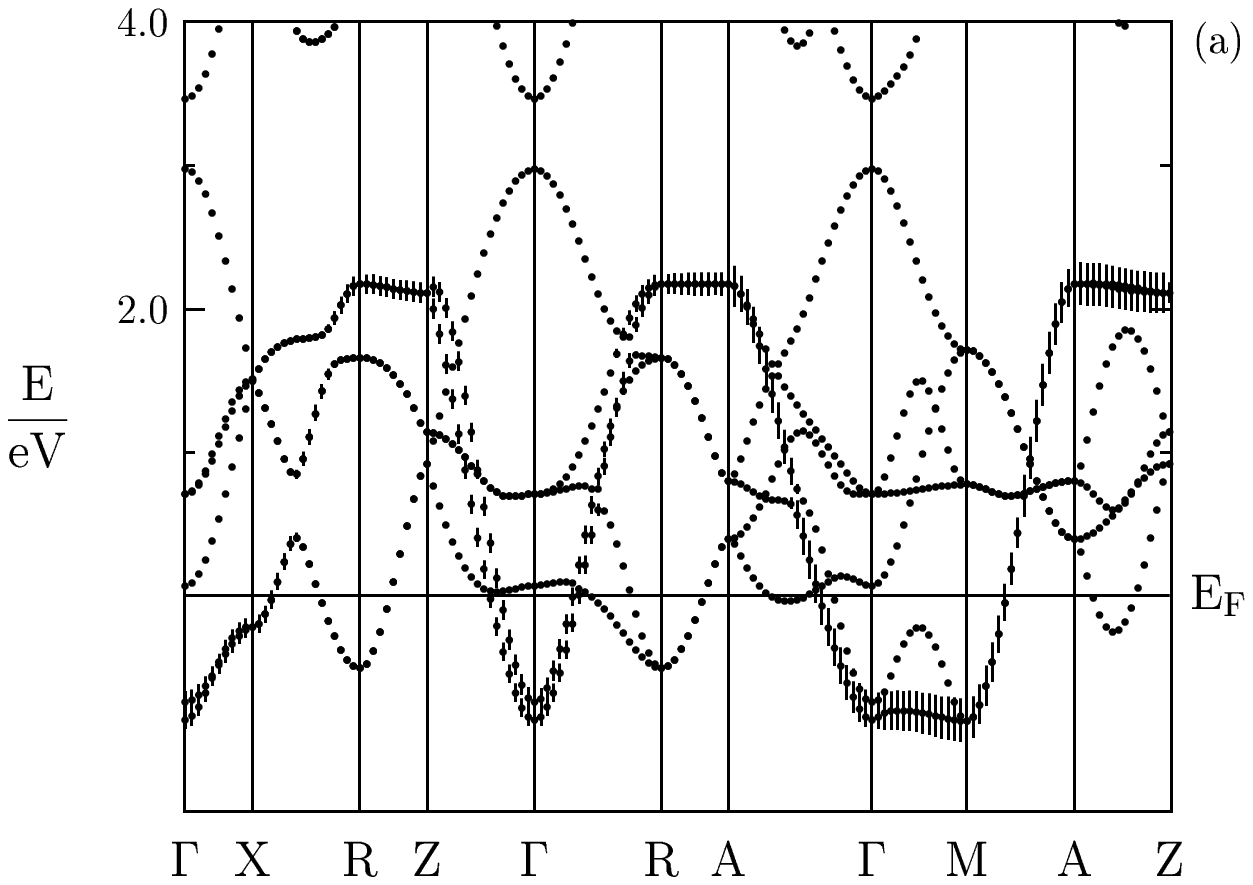}
    \includegraphics[width = 8.0cm]{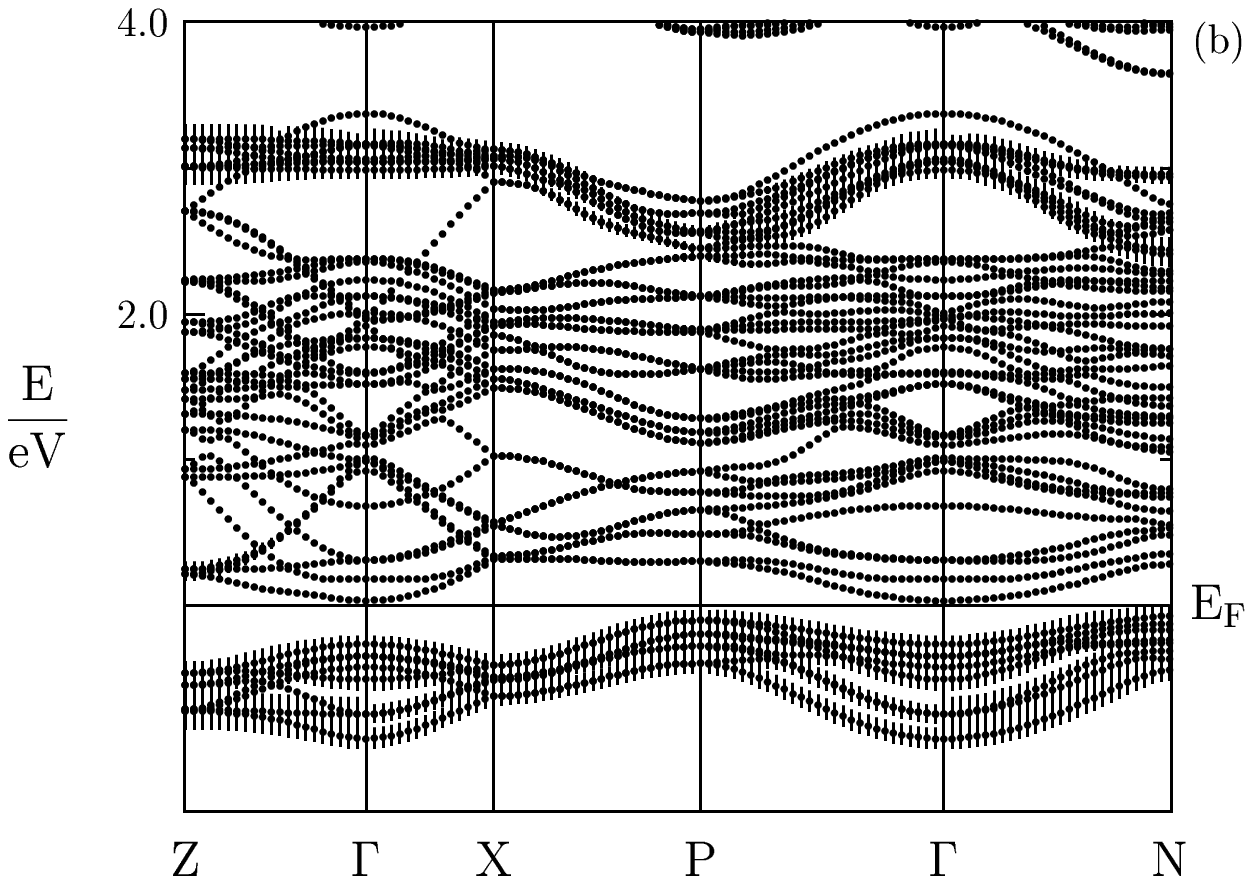}
    \caption{Weighted band structure of (a) metallic and (b) insulating 
             $ {\rm NbO_2} $ (see text).}
    \label{fig:bandsrut}
  \end{center}
\end{figure}

The results for both phases differ with respect to the energetical position  
of the O $ 2p $ group of bands as well as the shape of the $ t_{2g} $ 
partial DOS. In Fig.\ \ref{fig:dosrutt2g} 
these DOS are further separated into their symmetry components. To this 
end orbitals were transformed to a local coordinate system with the 
$ z $ and $ x $ axis parallel to the apical axis of the local octahedron 
and the rutile $ c $ axis, respectively \cite{habilmoo2pap}. While the 
$ d_{x^2-y^2} $ ($ \equiv d_{\parallel} $) and $ d_{yz} $ states mediate 
$ \sigma $-type $ d $-$ d $ overlap along the rutile $ c $ and $ a $ 
axes, respectively, the $ d_{xz} $ states account for $ \pi $-type 
overlap across the chains. Note that the latter two orbitals combine 
into the $ e_g^{\pi} $ states. The $ t_{2g} $ groups of bands are 
displayed in Fig.\ \ref{fig:bandsrut},  
where the $ d_{x^2-y^2} $ bands are highlighted by bars with widths 
proportional to the $ d_{x^2-y^2} $-contribution to the respective 
wave function. Thus, states with vanishing bars are of pure 
$ e_g^{\pi} $ character (apart from small $ e_g^{\sigma} $ as well 
as O $ 2p $ contributions). 

Most striking in Fig.\ \ref{fig:bandsrut}(a) is the nearly perfect 
one-dimensional dispersion of the $ d_{x^2-y^2} $ bands parallel to 
the rutile $ c $ axis, i.e.\ parallel to the niobium chains, which 
gives rise to the pronounced double peak structure in the partial DOS, 
Fig.\ \ref{fig:dosrutt2g}(a). In contrast, dispersion of the 
$ e_g^{\pi} $ bands is of the same order in all directions. This leads 
to the single broad structure of the $ d_{xz} $ partial DOS, while the 
$ d_{yz} $ states fall into two maxima due to the aforementioned inplane 
$ d $-$ d $ overlap. The DOS at $ E_F $ is comprised from similar 
contributions from all three $ t_{2g} $ orbitals explaining the rather 
isotropic electrical conductivity. Note that, except for the bands along 
the line X--R, the $ d_{\parallel} $ states hardly hybridize with the 
$ e_g^{\pi} $ states and thus coupling between both types of bands is 
mainly via charge conservation. For this reason, the $ d_{x^2-y^2} $ 
band may be regarded as a one-dimensional band in a three-dimensional 
embedding background of $ e_g^{\pi} $ bands. 

On going to insulating $ {\rm NbO_2} $ two distinct changes occur. 
i) The previously one-dimensional $ d_{x^2-y^2} $ band is split into 
bonding and antibonding branches due to Nb-Nb dimerization within 
the chains. As a consequence, the separation of the two peaks in the 
$ d_{x^2-y^2} $ partial DOS has increased by more than 1 eV and the 
contributions in between have been considerably reduced. 
ii) The $ e_g^{\pi} $ states experienced energetical upshift by 
$ \approx 0.5 $ eV due to increased $ p $-$ d $ overlap, causing almost 
complete depopulation of these orbitals. This is due to the much reduced 
Nb-O distances arising from the zigzag-like displacement of the Nb 
atoms. However, the lateral displacements also affect the inplane 
metal-metal bonding. For this reason, the two peak structure of the 
$ d_{yz} $ partial DOS has vanished and both $ e_g^{\pi} $ partial 
DOS have become more similar. 
Taken together these changes lead to complete separation of the low 
lying $ d_{x^2-y^2} $ bands from the $ e_g^{\pi} $ states, which gives 
rise to a band gap of 0.1 eV. This value was determined from checking 
band extrema on a $ 16 \times 16 \times 16 $ mesh within the Brillouin 
zone. Note that the hybridization between both types of bands is still 
negligible.

To conclude, the electronic structures of both metallic and insulating 
$ {\rm NbO_2} $ are well described within DFT and LDA.  
The near $ E_F $ electronic structure of metallic $ {\rm NbO_2} $ 
consists of two very weakly hybridizing types of $ t_{2g} $ bands, namely, 
$ d_{\parallel} $ states with one-dimensional dispersion parallel to 
the Nb chains and isotropically dispersing $ e_g^{\pi} $ bands. In 
the low-temperature structure Nb-Nb dimerization splits the $ 
d_{\parallel} $ band into bonding and antibonding branches, whereas 
the $ e_g^{\pi} $ states shift to higher energies due to reduced 
Nb-O distances. Since coupling between both types of bands is still by 
charge conservation rather than hybridization, the insulating state is 
interpreted as due to a Peierls instability of the $ d_{\parallel} $ bands 
in an embedding reservoir of $ e_g^{\pi} $ electrons. 

The results complement previous work on $ {\rm VO_2} $ as well as 
$ {\rm MoO_2} $ \cite{wentz94,habilmoo2pap} and give rise to a unified 
picture for the early transition metal dioxides, which explains 
(i) destabilization of the rutile structure and (ii) the metal-insulator 
transition of the $ d^1 $ members. Since the $ 4d $ bands are broader 
than the $ 3d $ bands metal-metal bonding is stronger in the fourth row 
dioxides this leading to the higher transition temperature of 
$ {\rm NbO_2} $ as compared to $ {\rm VO_2} $. In contrast, the absence 
of a gap in LDA calculations for the latter compound might be indicative 
of a considerable amount of electronic correlations. Here, a combination 
of LDA and dynamical mean field theory as recently successfully applied 
to $ {\rm V_2O_3} $ \cite{held01} might be more appropriate.

\begin{acknowledgments}
Valuable discussions with U.\ Eckern, K.-H.\ H\"ock, S.\ Horn, 
P.\ S.\ Riseborough, and D.\ Vollhardt are gratefully acknowledged. 
This work was supported by the Deutsche Forschungsgemeinschaft through 
Sonderforschungsbereich 484. 
\end{acknowledgments}

%
%


\begin{thebibliography}{99}

\bibitem{goode60}
J.\ B.\ Goodenough, 
Phys.\ Rev.\ {\bf 117}, 1442 (1960). 

\bibitem{adler68}
D.\ Adler, 
Rev.\ Mod.\ Phys.\ {\bf 40}, 714 (1968). 

\bibitem{goode71b}
J.\ B.\ Goodenough, 
in: {\em Progress in Solid State Chemistry},
    edited by H.\ Reiss (Pergamon Press, Oxford, 1971), Vol.\ 5, p.\ 145. 

\bibitem{zylbersztejn75}
A.\ Zylbersztejn and N.\ F.\ Mott, 
Phys.\ Rev.\ B {\bf 11}, 4383 (1975). 

\bibitem{wentz94}
R.\ M.\ Wentzcovitch, W.\ W.\ Schulz, and P.\ B.\ Allen, 
Phys.\ Rev.\ Lett.\ {\bf 72}, 3389 (1994); 
                    {\bf 73}, 3043 (1994). 


\bibitem{rice94}
T.\ M.\ Rice, H.\ Launois, and J.\ P.\ Pouget, 
Phys.\ Rev.\ Lett.\ {\bf 73}, 3042 (1994). 

\bibitem{imada98}
M.\ Imada, A.\ Fujimori, and Y.\ Tokura, 
Rev.\ Mod.\ Phys.\ {\bf 70}, 1039 (1998). 

\bibitem{andersson56longo70}
G.\ Andersson, 
Acta Chim.\ Scand.\ {\bf 10}, 623 (1956); 
J.\ M.\ Longo and P.\ Kierkegaard, 
Acta Chim.\ Scand.\ {\bf 24}, 420 (1970). 

\bibitem{janninck66}
R.\ F.\ Janninck and D.\ H.\ Whitmore, 
J.\ Phys.\ Chem.\ Solids {\bf 27}, 1183 (1966). 

\bibitem{sakata67}
T.\ Sakata, K.\ Sakata, and I.\ Nishida, 
phys.\ stat.\ sol.\ {\bf 20}, K155 (1967). 

\bibitem{marinder62}
B.-O.\ Marinder, 
Ark.\ Kemi {\bf 19}, 435 (1962). 

\bibitem{cheetham76}
A.\ K.\ Cheetham and C.\ N.\ R.\ Rao, 
Acta Cryst.\ B {\bf 32}, 1579 (1976). 

\bibitem{pynn76b}
R.\ Pynn, J.\ D.\ Axe, and R.\ Thomas, 
Phys.\ Rev.\ B {\bf 13}, 2965 (1976). 

\bibitem{rogers69mattheiss76}
D.\ B.\ Rogers, R.\ D.\ Shannon, A.\ W.\ Sleight, and J.\ L.\ Gillson, 
Inorg.\ Chem.\ {\bf 8}, 841 (1969); 
L.\ F.\ Mattheiss, 
Phys.\ Rev.\ B {\bf 13}, 2433 (1976). 

\bibitem{posternak79}
M.\ Posternak, A.\ J.\ Freeman, and D.\ E.\ Ellis, 
Phys.\ Rev.\ B {\bf 19}, 6555 (1979). 

\bibitem{sasaki81}
T.\ A.\ Sasaki and T.\ Soga, 
Physica B {\bf 111}, 304 (1981). 

\bibitem{habilmoo2pap}
V.\ Eyert, 
Habilitation thesis, Universit\"at Augsburg, 1998; 
V.\ Eyert, R.\ Horny, K.-H.\ H\"ock, and S.\ Horn,  
J.\ Phys.: Cond.\ Matt.\ {\bf 12}, 4923 (2000). 

\bibitem{brauer41}
G.\ Brauer,
Z.\ Anorg.\ Allg.\ Chem.\ {\bf 248}, 1 (1941). 

\bibitem{ruedorff64}
W.\ R\"udorff and H.-H.\ Luginsland, 
Z.\ Anorg.\ Allg.\ Chem.\ {\bf 334}, 125 (1964). 

\bibitem{sakata69}
K.\ Sakata, 
J.\ Phys.\ Soc.\ Japan {\bf 26}, 582 (1969); 
                       {\bf 26}, 867 (1969); 
                       {\bf 26}, 1067 (1969). 

\bibitem{sakata69d}
K.\ Sakata, I.\ Nishida, M.\ Matsushima, and T.\ Sakata, 
J.\ Phys.\ Soc.\ Japan {\bf 27}, 506 (1969). 

\bibitem{bolzan94}
A.\ A.\ Bolzan, C.\ Fong, B.\ J.\ Kennedy, and C.\ J.\ Howard, 
J.\ Solid State Chem.\ {\bf 113}, 9 (1994); 
Acta Cryst.\ B {\bf 53}, 373 (1997). 

\bibitem{shapiro74}
S.\ M.\ Shapiro, J.\ D.\ Axe, G.\ Shirane, and P.\ M.\ Raccah, 
Solid State Commun.\ {\bf 15}, 377 (1974). 

\bibitem{pynn78}
R.\ Pynn, J.\ D.\ Axe, and P.\ M.\ Raccah, 
Phys.\ Rev.\ B {\bf 17}, 2196 (1978). 

\bibitem{gervais85}
F.\ Gervais and W.\ Kress, 
Phys.\ Rev.\ B {\bf 31}, 4809 (1985). 

\bibitem{beatham79}
N.\ Beatham and A.\ F.\ Orchard, 
J.\ El.\ Spec.\ Rel.\ Phen.\ {\bf 16}, 77 (1979). 

\bibitem{umrigar80}
C.\ Umrigar and D.\ E.\ Ellis, 
Phys.\ Rev.\ B {\bf 21}, 852 (1980). 

\bibitem{xu89}
J.\ H.\ Xu, T.\ Jarlborg, and A.\ J.\ Freeman, 
Phys.\ Rev.\ B {\bf 40}, 7939 (1989). 

\bibitem{wkgrevasw}
A.\ R.\ Williams, J.\ K\"ubler, and C.\ D.\ Gelatt, Jr.,  
Phys.\ Rev.\ B {\bf 19}, 6094 (1979); 
V.\ Eyert, 
Int.\ J.\ Quantum Chem.\, {\bf 77}, 1007 (2000).

\bibitem{vpop}
V.\ Eyert and K.-H.\ H\"ock, 
Phys.\ Rev.\ B {\bf 57}, 12727 (1998). 

\bibitem{methfessel89}
M.\ S.\ Methfessel and A.\ T.\ Paxton, 
Phys.\ Rev.\ B {\bf 40}, 3616 (1989).

\bibitem{held01}
K.\ Held, G.\ Keller, V.\ Eyert, D.\ Vollhardt, and V.\ Anisimov, 
Phys.\ Rev.\ Lett.\ {\bf 86}, 5345 (2001). 

\end{thebibliography}
\end{document}